\documentclass[11pt]{article}

\usepackage{comment, url}

\def\nottoobig#1{{\hbox{$\left#1\vcenter to1.111\ht\strutbox{}\right.\n@space$}}}

\newtheorem{remark}{Remark}
\newtheorem{theorem}{Theorem}[section]
\newtheorem{corollary}[theorem]{Corollary}
\newtheorem{lemma}[theorem]{Lemma}
\newtheorem{fact}[theorem]{Fact}
\newtheorem{claim}[theorem]{Claim}
\newtheorem{proposition}[theorem]{Proposition}

\newtheorem{definition}[theorem]{Definition}

\usepackage{geometry}                
\usepackage{graphicx}
\usepackage{amssymb}
\usepackage{graphicx}
\usepackage{amssymb}
\usepackage{amsmath}
\newcommand{\dom}{{\rm dom}}

\newtheorem{thm}{Theorem}

\newtheorem{sch}[thm]{Scholium}
\usepackage{amsmath}
\newcommand{\ie}{$\mbox{i.e.}$}

\renewcommand{\lll}{\lVert}
\newcommand{\rrr}{\rVert}

\newcommand{\nat}{\mathbb{N}}
\newcommand{\real}{\mathbb{R}}

\newcommand{\rest}{{\upharpoonright}}
\newcommand{\bin}{\mbox{\rm  bin}}

\if01
\fi

\makeatletter
\def\@listI{\leftmargin\leftmargini \parsep 4.5pt plus 1pt minus 1pt\topsep6pt plus 2pt minus 2pt \itemsep  2pt plus 2pt minus 1pt}

\let\@listi\@listI
\@listi
\makeatother
\def\QED{$\hspace*{\fill}\Box$}
\author{{Cristian S. Calude}
\thanks{Department of Computer Science, University of Auckland, New Zealand, {\tt www.cs.auckland.ac.nz/}  {\tt \~{ }cristian}. Calude was supported in part by UARC Grant 3607894/9343 and CS-PBRF Grant.}
\and 
{Marius Zimand}
\thanks{ Department of Computer and Information Sciences, Towson University,
Baltimore, MD, USA, {\tt  http://triton.towson.edu/\~{ }mzimand}.  Zimand was supported by NSF grant CCF 0634830. Part of this work was done while visiting the CDMTCS of the University of Auckland, New Zealand.}}
\date{}

\title{{\bf Algorithmically Independent Sequences}}

\begin{document}

\maketitle
\begin{abstract} 
Two objects are independent if they do not affect each other.  Independence 
is well-understood in classical information theory, but less in algorithmic information theory.
Working in the framework of  algorithmic information theory, the paper proposes two types of independence for arbitrary infinite binary sequences and studies their properties.  Our two proposed notions of independence have some of the intuitive properties that one naturally expects. 
For example, for every sequence $x$, the set of sequences that are
 independent (in the weaker of the two senses) with $x$ has measure one.
 For both notions of independence we  investigate to what extent pairs of independent  sequences, can be effectively constructed via Turing reductions (from one or more input sequences). In this respect, we prove several impossibility results. 
 For example, it is shown that there is no effective way of producing from an arbitrary sequence with positive constructive Hausdorff dimension two sequences that are independent (even in the weaker type of independence) and have super-logarithmic complexity.  Finally, a few conjectures and open questions are discussed.
\end{abstract}
\section{Introduction}  

Intuitively, two objects are independent if they do not affect each other.  The concept is well-understood in classical information theory. There, the objects  are random variables, the information in a random variable is its Shannon entropy, and two random variables $X$ and $Y$ are declared to be independent if the information in the join $(X,Y)$ is equal to the sum of the information in $X$ and the information in $Y$. This is equivalent to saying that the information in $X$ conditioned by $Y$ is equal to the information in $X$, with the interpretation that, on average, knowing a particular value of $Y$ does not affect the information in $X$.

The notion of independence has been defined in algorithmic information theory as well for finite strings~\cite{Cha82}. The approach is very similar. This time the information in a string $x$ is the  complexity (plain or prefix-free) of $x$, and two strings $x$ and $y$ are independent if the information in the join string $\langle x, y \rangle$ is equal to the sum of the information in $x$ and the information in $y$, up to logarithmic (or, in some cases, constant) precision.

The case of infinite sequences (in short, sequences) has been less studied. An inspection of the literature reveals that for  this setting, independence has been considered to be synonymous with pairwise relative randomness, $\mbox{i.e.}$, two sequences $x$ and $y$ are said to be independent if they are (Martin-L\"{o}f) random relative to each other (see~\cite{vlam:j:randomness, dow-hir:b:algrandom}). The effect of this approach is that the notion of independence is confined to the situation where the sequences are random.

The main objective of this paper is to put forward a concept of independence that applies to \emph {all} sequences. One can envision various ways for doing this. One possibility is to use Levin's notion of mutual information for sequences~\cite{lev:j:mutualinformation} (see also the survey paper~\cite{gru-vit:t:shankolm}) and declare two sequences to be independent if their mutual information is small. If one pursues this direction, the main issue is to determine the right definition for ``small." We take another approach, which consists in extending in the natural way the notion of independence from finite strings to sequences. This leads us to two concepts: \emph {independence} and \emph{finitary-independence}. We say that (1) two sequences $x$ and $y$ are independent if, for all $n$, the complexity of $x \rest n$ (the prefix of $x$ of length $n$)  and the complexity of $x \rest n$ relativized with $y$ are within $O(\log n)$ (and the same relation holds if we swap the roles of $x$ and $y$), and (2) two sequences $x$ and $y$ are finitary-independent if, for all $n$ and $m$, the complexity of $x \rest n$  and the complexity of $x \rest n$ given $y\rest m$ are within $O(\log n + \log m)$ (and the same relation holds if we swap the roles of $x$ and $y$).
We have settled for the additive logarithmical term of precision (rather than some higher accuracy) since this provides robustness with respect to the type of complexity (plain or prefix-free) and other technical advantages.

We establish a series of basic facts regarding the proposed notions of independence. We show that independence is strictly stronger than finitary-independence. The two notions of independence apply to a larger category of sequences than the family of random sequences, as intended. However, they are too rough for being relevant for computable sequences. It is not hard to see that a computable sequence $x$ is independent with any other sequence $y$, simply because the information in $x$ can be obtained directly. In fact, this type of trivial independence holds for a larger type of sequences, namely for any $H$-trivial sequence, and trivial finitary-independence holds for any sequence $x$ whose prefixes have logarithmic complexity. It seems that for this type of sequences (computable or with very low complexity) a more refined definition of independence is needed (perhaps, based on resource-bounded  complexity). We show that the two proposed notions of independence have some of the intuitive properties that one naturally expects. For example, for every sequence $x$, the set of sequences that are
 finitary-independent with $x$ has measure one. The same issue for independence remains open.

We next investigate to what extent pairs of independent, or finitary-independent sequences, can be effectively constructed via Turing reductions. For example, is there a Turing reduction $f$ that given oracle access to an arbitrary sequence $x$ produces a sequence that is finitary-independent with $x$?
Clearly, if we allow the output of $f$ to be a computable sequence, then the answer is positive by the type of trivial finitary-independence that we have noted above. We show that if we insist that the output of $f$ has super-logarithmic complexity whenever $x$ has positive constructive Hausdorff dimension, then the answer is negative. In the same vein, it is shown that there is no effective way of producing from an arbitrary sequence $x$ with positive constructive Hausdorff dimension two sequences that are finitary-independent and have super-logarithmic complexity. 

Similar questions are considered for the situation when we are given two (finitary-) independent sequences. It is shown that there are independent sequences $x$ and $y$ and a Turing reduction $g$ such that $x$ and $g(y)$ are not independent. This appears to be a bad artifact of the notion of independence proposed in this paper. We consider that this is the only counter-intuitive effect of our definitions. We do not know if a similar phenomenon holds for finitary-independence. On the other hand, for any independent sequences $x$ and $y$ and for any Turing reduction $g$, $x$ and $g(y)$ are finitary-independent. 

We also raise the question on whether given as input several (finitary-) independent sequences $x$ and $y$ it is possible to effectively build a new sequence that is (finitary-) independent (not in the trivial way) with each sequence in the input. It is observed that the answer is positive if the sequences in the input are random, but for other types of sequences the question remains open. The same issue can be raised regarding finite strings and for this case  a positive answer is obtained. Namely, it is shown that given three independent finite strings $x$, $y$ and $z$ with linear complexity, one can effectively construct a new string that is independent with each of $x, y$ and $z$, and has high complexity and length a constant fraction of the length of $x, y$ and $z$.

\subsection{Preliminaries}
${\mathbb N}$ denotes the set of non-negative integers; the size of a finite set $A$ is denoted $|| A||$. Unless stated otherwise, all numbers are in ${\mathbb N}$ and all logs are in base 2.
We work over the binary alphabet $\{0,1\}$. A string is an element of $\{0,1\}^*$ and a sequence is an element of $\{0,1\}^{\infty}$. If $x$ is a string, $|x|$ denotes its length; $xy$ denotes the concatenation of the strings $x$ and $y$. If $x$ is a string or a sequence, $x(i)$ denotes the $i$-th bit of $x$ and  $x\rest n$ is the substring $x(1) x(2) \cdots x(n)$. For two sequences $x$ and $y$, $x \oplus y$ denotes the sequence $x(1) y(1) x(2) y(2) x(3) y(3) \cdots $ and $x ~{\mathrm{XOR}}~ y$ denotes the sequence $(x(1) ~{\mathrm{XOR}}~ y(1)) (x(2) ~{\mathrm{XOR}}~ y(2)) (x(3) ~{\mathrm{XOR}}~ y(3)) \cdots$, where $(x(i) ~{\mathrm{XOR}}~ y(i))$ is the sum modulo $2$ of the bits $x(i)$ and $y(i)$. We identify a sequence $x$ with the set $\{n \in \nat \mid x(n) = 1\}$.  We say that a sequence $x$ is computable (computably enumerable, or c.e.) if the corresponding set is computable (respectively, computably enumerable, or c.e.).  If $x$ is c.e., then for  every  
 $s \in \nat$, $x_s$ is the sequence corresponding to the set of elements enumerated within $s$ steps by some machine $M$  that enumerates $x$ (the machine $M$ is given in the context). We also identify a sequence $x$ with the real number in the interval $[0,1]$ whose binary writing is $0.x(1) x(2) \cdots$. A sequence $x$ is said to be left c.e. if the corresponding real number $x$ is the limit of a computable increasing sequence of rational numbers. The plain and the prefix-free complexities of a string are defined in the standard way; however we need to provide a few details regarding the computational models. The machines that we consider process information given in three forms: (1) the input, (2) the oracle set, (3) the conditional string. Correspondingly, a universal machine has 3 tapes:
\begin{itemize}
	\item one tape for the input and work,
	\item one tape for storing the conditional string,
	\item one tape (called the oracle-query tape) for formulating queries to the oracle.
\end{itemize}
The oracle is a  string or a  sequence. If the machine enters the query state and the value written in binary on the oracle-query tape is $n$, then the machine gets the $n$-th bit in the oracle, or if $n$ is larger than the length of the oracle, the machine enters an infinite loop. 

We fix such a universal machine $U$. The notation $U^w(u \mid v)$ means that the input is $u$, the conditional string $v$ and the oracle  is given by $w$, which is a string or a sequence. The plain complexity of a string $x$ given the oracle $w$ and the conditional string $v$ is $C^w(x \mid v) = \min\{|u| \mid U^w(u \mid v) = x \}$. There exists a constant $c$ such that for every $x, v$ and $w$ $C^w(x \mid v) < |x| + c$.

A machine is prefix-free (self-delimiting)  if its domain is a prefix-free set. There exist universal prefix-free machines; we fix such a machine $U$; the prefix-free complexity of a string $x$ given the oracle $w$ and the conditional string $v$ is $H^w(x \mid v) = \min\{|u| \mid U^w(u \mid v) = x \}$.  

In case $w$ or $v$ are the empty strings, we omit them in  $C(\cdot)$ and $H(\cdot)$. Throughout this paper we use the $O(\cdot)$ notation to hide constants that depend only on the choice of the universal machine underlying the definitions of the complexities $C$ and $H$. Since the prefix-free universal machine is a particular type of machine, it follows that $C^w(x \mid v) < H^w(x \mid v) + O(1)$, for every $x, v$ and $w$. The reverse inequality between $C(\cdot)$ and $H(\cdot)$ also holds true, within an additive logarithmic term, and is obtained as follows.
For example, a string $x = x(1) x(2) \cdots x(n)$ can be coded in a self-delimiting way by $x \mapsto code(x)= \underbrace{11 \cdots 1}_{|\bin(n)|} 0 \bin(n) x(1) x(2) \cdots x(n),$ where $\bin(
n)$ is the binary representation of $n \in \nat$. Note that $|code(x)| = |x| + 2 \log |x| + O(1)$. This implies that for every $x,v$, and $w$, 
\begin{equation}
\label{e:plain-prefix}
C^w(x \mid v) > H^w(x \mid v) - 2 \log|x| - O(1).
\end{equation} 
The following inequalities hold for all strings $x$ and $y$:
\begin{equation}
\label{e:lifting}
C^y(x) \leq C(x \mid y) + 2 \log |y| + O(1),
\end{equation}
\begin{equation}
\label{e:symmetry}
| C(xy) - (C(x|y) + C(y)) | \leq O(\log C(x) + \log C(y)).
\end{equation}
The first inequality is easy to derive directly; the second one is called the Symmetry of Information Theorem, see \cite{zvo-lev:j:kol}.

There are various equivalent definitions for (algorithmic) random sequences as defined by Martin-L\"of \cite{mar-lof:j:mltests}  (see \cite{Cris}). In what follows we will use
the (weak) complexity-theoretic one \cite{cha:j:ait} using  the prefix-free complexity: A  sequence $x$ is Martin-L\"{o}f random (in short, random) if there is a constant $c$ such that for every $n$, $H(x \rest n) \geq n-c$. The set of random sequences has constructive (Lebesgue) measure one \cite{mar-lof:j:mltests}.

The sequence $x$ is random relative to the sequence $y$ if there is a constant $c$ such that for every $n$, $H^y(x \rest n) \geq n-c$. Note that if $x$ is random, then for every $n$, $C(x \rest n) \geq n - 2\log n -O(1)$ (by inequality (\ref{e:plain-prefix})). A similar inequality also holds for the relativized complexities, i.e. for all  $ x$ that are random relative to $y$ and for all $n$, 
 $C^y(x \rest n ) > n - 2 \log n - O(1)$.  These results will be repeatedly used throughout  the paper.

In \cite{vlam:j:randomness} van Lambalgen proves  that $x \oplus y$ is random iff $x$ is random and $y$ is random relative to $x$. This implies that if $x$ is random and $y$ is random relative to $x$ then $x$ is random relative to $y$.

The constructive Hausdorff dimension of a sequence $x$---which is the direct effectivization of  ``classical Hausdorff dimension''---defined by
$\mathrm{dim}(x)=\lim \inf_{n \rightarrow \infty} C(x \rest n)/n\left( = \lim \inf_{n \rightarrow \infty} H(x \rest n)/n\right)$, 
 measures intermediate levels of randomness
 (see \cite{rya:j:dimension, sta:93,   tad:j:partialrand, may:j:dimension-kol, lut:j:dimension,    rei:t:thesis,sta:j:dimension, cal-sta-ter:j:partialrand, dow-hir-nie-ter:j:calibrating}).  

\if01Originally it has been defined using measure-theoretical tools but we give an alternative equivalent definition (for the proof and history of this equivalence see~\cite{may:j:dimension-kol, rya:j:dimension, sta:j:dimension} ) that involves plain (or prefix-free) complexity. Thus, the constructive Hausdorff dimension of a sequence $x$, denoted $\mathrm{dim}(x)$ is $\lim \inf_{n \rightarrow \infty} \frac{C(x \rest n)}{n} $ ($ = \lim \inf_{n \rightarrow \infty} \frac{H(x \rest n)}{n}$). 
\fi

A Turing reduction $f$ is an oracle Turing machine; $f(x)$ is the language computed by $f$ with oracle $x$, assuming that $f$ halts on all inputs when working with oracle $x$ (otherwise we say that $f(x)$ does not exist). In other words, if $n \in f(x)$ then the machine $f$ on input $n$ and with oracle $x$ halts and outputs $1$ and if  $n \not\in f(x)$ then the machine $f$ on input $n$ and with oracle $x$ halts and outputs $0$. The function \emph{use} is defined as follows: $use_f^x(n)$ is the index of the rightmost position on the tape of $f$ accessed during the computation of $f$ with oracle $x$ on input $n$. The Turing reduction $f$ is a \emph{wtt-reduction} if there is a computable function $q$ such that $use_f^x(n) \leq q(n)$, for all $n$. The Turing reduction $f$ is a \emph{truth-table reduction} if $f$ halts on all inputs for every oracle. A truth-table reduction is a wtt-reduction.

\section{Defining independence}

The basic idea is to declare  that two objects are independent if none of them contains significant information about the other one. Thus, if in some formalization, $I(x)$ denotes the information in $x$ and $I(x \mid y)$ denotes the information in $x$ given $y$, $x$ and $y$ are independent if $I(x) - I(x \mid y)$ and $I(y) - I(y \mid x)$ are both small. In this paper we work in the framework of algorithmic information theory. In this setting, in case $x$ is a  string, $I(x)$ is the complexity of $x$ (where for the ``complexity of $x$" there are several possibilities, the main ones being the plain complexity or the prefix-free complexity). 

The independence of  strings was studied in~\cite{Cha82}: two  strings are independent  if
$I(xy) \approx I(x) + I(y)$. This approach  motivates our  Definition~\ref{d:strongindep} and Definition~\ref{d:weakindep}.

In case $x$ is an infinite sequence, the information in $x$ is characterized by the sequence $(I(x \rest n))_{n \in \nat}$ of information in the initial segments of $x$. In the infinite case, for the information upon which we condition ($\mbox{e.g.}$, the $y$ in $I(x \mid y)$), there are two possibilities: either the entire sequence is available in the form of an oracle, or we consider initial segments of it. Accordingly, we propose two notions of independence.

 \begin{definition} {\rm(}{\bf The ``integral" type of independence}{\rm )}
 \label{d:strongindep}
 Two sequences $x$ and $y$ are {\rm independent} if $C^x(y\rest n) \geq C(y\rest n) - O(\log n )$ and $C^y(x\rest n) \geq C(x\rest n) - O(\log n)$. 
 \end{definition}
  
\medskip

\begin{definition}  {\rm(}{\bf The finitary type of independence}{\rm )}
\label{d:weakindep}
Two sequences $x, y$ are {\rm  finitary-independent} if for all natural numbers $n$ and $m$,
\[
C(x \rest n ~y \rest m) \geq C(x\rest n) + C(y\rest m) - O(\log(n) + \log(m)).
\]
\end{definition}

\begin{remark}
{\rm 
We will show in Proposition~\ref{p:cond}, that the inequality in Definition~\ref{d:weakindep} is equivalent to saying that for all $n$ and $m$, $C(x\rest n \mid y\rest m) \geq C(x\rest n) - O(\log n + \log m)$, which is the finite analogue of the property in Definition~\ref{d:strongindep} and is in line with our discussion above.
} 
\end{remark}

\begin{remark}
\label{r:indepimplieswindep}
{\rm 
If $x$ and $y$ are independent, then they are also finitary-independent (Proposition~\ref{p:indepimplieswindep}). The converse is not true (Corollary~\ref{c:windepnotindep}).
} 
\end{remark}

\begin{remark}
\label{r:plain-prefix}
{\rm 
The proposed definitions use the plain complexity $C(\cdot)$, but we could have used the prefix-free complexity as well, because the two types of complexity are within an additive logarithmic term. Also, in Definition~\ref{d:weakindep} (and throughout this paper), we use concatenation to represent the joining of two strings. However, since any reasonable pairing function $\langle x, y \rangle$ satisfies $| ~|\langle x, y \rangle| - |xy| ~| < O(\log|x| + \log|y|)$, it follows that $|C(<x,y>) - C(xy)| < O(\log|x| + \log|y|)$, and thus any reasonable pairing function could have been used instead.
} 
 \end{remark}
 
\begin{remark} 
{\rm 
A debatable issue is the subtraction of the logarithmic term. Indeed, there are other natural possibilities. We argue that our choice has certain advantages over other possibilities that come to mind. 

Let us focus on the definition of finitary-independence. We want $C(x\rest n ~y\rest m) \geq C(x\rest n) + C(y\rest n) - O(f(x) + f(y))$, for all $n, m$, where $f$ should be some ``small" function.
We would like  the following two properties to hold:
\begin{itemize}

\item[(A)] the sequences $x$ and $y$ are finitary-independent iff  $C(x\rest n \mid y\rest m) > C(x\rest n) - O(f(x\rest n) + f(y\rest m))$, for all $n$ and $m$,

\item[(B)] if $x$ is ``somewhat'' random and $y = 0^{\omega}$, then $x$ and $y$ are finitary-independent.
\end{itemize}

Other natural possibilities for the definition could be:

(i) if $f(x) = C(|x|)$, the definition of  finitary independence--(i) would now be:

\[
C(x\rest n ~y\rest m) \geq C(x\rest n) + C(y\rest m) - O(C(n) + C(m)),
\]

or  

(ii) if $f(x) = \log C(x)$,  the definition of finitary-independence--(ii) would now be:

\[
C(x\rest n ~y\rest m) \geq C(x\rest n) + C(y\rest m) - O(\log C(x\rest n) + \log C(y\rest m)).
\]

If sequences $x$ and $y$ satisfy (i), or (ii), then they  also satisfy  Definition~\ref{d:weakindep}.

Variant (i) implies (B), but not(A) (for example, consider sequences $x$ and $y$ with $C(n) << \log C(x \rest n)$ and $C(m) << \log C(y\rest m)$, for infinitely many $n$ and $m$). Variant (ii) implies (A), but does not imply (B) (for example if for infinitely many $n$, $C(x\rest n) = O(\log^3 n)$;  take such a value $n$, let $p$ be a shortest description of $x \rest n$, and let $m$ be the integer whose binary representation is $1p$. Then $x\rest n$ and $0^\omega \rest m$, do not satisfy (B)).
The proposed definition implies both (A) and (B).

Another advantage is the robustness properties from  Remark~\ref{r:plain-prefix}.
}
\end{remark}

 \begin{remark}
 \label{r:trivial}
 {\rm If the sequence $x$ is computable, then $x$ is independent with every sequence $y$. In fact a stronger fact holds. A sequence is called $H$-trivial if, for all $n$, $H(x \rest n) \leq H(n) + O(1)$. This is a notion that has been intensively studied recently (see \cite{dow-hir-nie-ter:j:calibrating}). Clearly every computable sequence is $H$-trivial, but the converse does not hold~\cite{zam:t:kolmog, sol:t:kolmog}.  If $x$ is $H$-trivial, then it is independent with every sequence $y$. Indeed, $H^y(x \rest n) \geq H(x\rest n) - O(\log n)$, because $H(x \rest n) \leq H(n) + O(1) \leq \log n + O(1)$, and $H^x(y \rest n) \geq H(y \rest n) - O(\log n)$, because, in fact, 
 $H^x(y \rest n)$ and $H(y \rest n)$ are within a constant of each other~\cite{nie:j:ktrivial}. The same inequalities hold if we use the $C(\cdot)$ complexity (see Remark~\ref{r:plain-prefix}).
 
 For the case of finitary-independence, a similar phenomenon holds for a (seemingly) even larger class.
 \begin{definition} A sequence $x$ is called C-logarithmic if $C(x\rest n) = O(\log n)$.
\end{definition}
It can be shown (for example using Proposition~\ref{p:cond}, (a)) that if $x$ is C-logarithmic, then it is finitary-independent with every sequence $y$.

Note that every  
 sequence $x$ that is the characteristic sequence of a $\mbox{c.e.}$ set is C-logarithmic. This follows from the observation that, for every  
 $n$, the initial segment $x \rest n$ can be constructed given the number of $1$'s in $x \rest n$ (an information which can be written with $\log n$ bits) and the finite description of the enumerator of the set represented by $x$.
If a sequence is $H$-trivial then it is C-logarithmic, but the converse probably does not hold.

In brief, the notions of independence and finitary-independence are relevant for strings having complexity above that of $H$-trivial sequences, respectively C-logarithmic sequences. The cases of independent (finitary-independent) pairs $(x,y)$, where at least one of $x$ and $y$ is $H$-trivial (respectively, C-logarithmic) will be referred to as \emph{trivial independence}.
 } 
 \end{remark}
 
\begin{remark}
\label{r:properties}
{\rm 
Some desirable properties of the independence relation are:
\begin{itemize}

	\item[P1.] Symmetry: $x$ is independent with $y$ iff  $y$ is independent with $x$.
	\item[P2.] Robustness under type of complexity (plain or prefix-free).
	\item[P3.] If $f$ is a Turing reduction, except for some special cases, $x$ and $f(x)$ are dependent (``independence cannot be created'').
	\item[P4.] For every  
 $x$, the set of sequences that are dependent with $x$ is small (\ie, it has measure zero).
	
	\end{itemize}
Clearly both the independence and the finitary-independence relations satisfy P1. They also satisfy P2, as we noted in Remark~\ref{r:plain-prefix}. 


It is easy to see that the independence relation satisfies P3, whenever we require that the initial segments of $x$ and $f(x)$ have plain complexity $\omega (\log n)$ (because $C^x(f(x) \rest n) = O(\log n)$, while $C(f(x) \rest n) = \omega(\log n)$).
We shall see that the finitary-independence relation satisfies P3 under some stronger assumptions for $f$ and $f(x)$  (see Section~\ref{s:onesource} and in particular Theorem~\ref{t:wtt-1}). 

We do not know whether the independence relation satisfies P4. Theorem~\ref{t:measone} shows that the finitary-independence relation satisfies P4.
} 
\end{remark}

 \subsection{Properties of independent and finitary-independent sequences}
 \label{s:wealindep}
The following simple properties of finitary-independent sequences are technically useful in some of the next proofs.

\begin{proposition}
\label{p:cond}
\begin{itemize}
\item[\rm (a)] Two sequences $x$ and $y$ are finitary-independent $\Leftrightarrow$ for all $n$ and $m$, $C(x\rest n \mid y\rest m) \geq C(x\rest n) - O(\log n + \log m)$.
\item [\rm (b)] Two sequences $x$ and $y$ are finitary-independent if and only if for all $n$,
$C(x\rest n ~y\rest n) \geq C(x\rest n) + C(y\rest n) - O(\log(n))$.
\item [\rm (c)] Two sequences $x$ and $y$ are finitary-independent if and only if for all $n$,
$C(x\rest n \mid y\rest n) \geq C(x\rest n) - O(\log(n))$.
\item [\rm (d)] If $x$ and $y$ are not finitary-independent, then for every  
 constant $c$ there are infinitely many $n$ such that 
$C(x\rest n ~y\rest n) < C(x \rest n) + C(y \rest n) - c \log n$.
\item [\rm (e)] If $x$ and $y$ are not finitary-independent, then for every  
 constant $c$ there are infinitely many $n$ such that 
$C(x\rest n \mid y\rest n) < C(x \rest n) - c \log n$.
\end{itemize}
\end{proposition}

{\em Proof}.
 We use the following inequalities which hold  for every  
 strings $x$ and $y$ (they follow from the Symmetry of Information Equation~(\ref{e:symmetry})):
\begin{equation}
\label{e:condgeq}
C(xy) \geq C(x) + C(y \mid x) - O(\log|x| + \log |y|),
\end{equation}


and
\begin{equation}
\label{e:condleq}
C(xy) \leq C(x) + C(y \mid x) + O(\log|x| + \log |y|).
\end{equation}
\phantom{xx.} 
(a)``$\Rightarrow$"
\begin{equation*}
\begin{array}{ll}
\quad\quad\quad  C(x\rest n \mid y\rest m) & \geq C(x\rest n ~y\rest m) - C(y\rest m) - O(\log n + \log m) \quad \quad (\mbox{by~(\ref{e:condleq})}) \\
& \geq C(x\rest n) + C(y\rest m) - C(y\rest m) - O(\log n + \log m) \quad \quad (\mbox{by independence}) \\
& = C(x\rest n) - O(\log n + \log m).
\end{array}
\end{equation*}
\phantom{xx. (a)}``$ \Leftarrow$" 
\begin{equation*}
\begin{array}{ll}
C(x\rest n ~y\rest m) & \geq C(y\rest m) + C(x\rest n \mid y\rest m) - O(\log n + \log m) \quad \quad (\mbox{by~(\ref{e:condgeq})}) \\
& \geq C(y\rest m) + C(x\rest n) - O(\log n + \log m) \quad \quad (\mbox{by hypothesis}).
\end{array}
\end{equation*}
\phantom{xx.} (b)  ``$\Rightarrow$"
 Take $n=m$.

\phantom{xx. (b)\quad}``$\Leftarrow$" Suppose $n \geq m$ (the other case can be handled similarly).
\begin{equation*}
\begin{array}{ll}
C(x\rest n ~y\rest m) & \geq C(y\rest m) + C(x\rest n \mid y\rest m) - O(\log(n) + \log(m)) \quad \quad (\mbox{by~(\ref{e:condgeq})}) \\
& \geq C(y\rest m) + C(x\rest n \mid y\rest n) - O(\log(n) + \log(m)) \\
& \geq C(y\rest m) + C(x\rest n) - O(\log(n) + \log(m)) \quad \quad (\mbox{by (a)}).
 \end{array}
\end{equation*}
\phantom{xx.} (c) This follows from (b) with a similar  proof as for   (a).
\smallskip

(d) Suppose that for some constant $c$ the inequality holds only for finitely many $n$. Then one can choose a constant $c' > c$ for which the opposite inequality holds for every  
 $n$, which by (b) would imply the finitary-independence of $x$ and $y$.
\smallskip

(e) Follows from (c), in a similar way  as (d) follows from (b).~\QED

\medskip

\begin{proposition}
\label{p:indepimplieswindep}
If the sequences $x$ and $y$ are independent, then they are also finitary-independent.
\end{proposition}
{\em Proof}.
 Suppose $x$ and $y$ are not finitary-independent. By Proposition~\ref{p:cond} (e), for every  
 constant $c$ there are infinitely many $n$ such that $C(x\rest n \mid y \rest n) < C(x \rest n)- c \cdot \log n$. Taking into account inequality~(\ref{e:lifting}), we obtain $C^y(x\rest n) < C(x \rest n) - (c-3) \log n$, for infinitely many $n$, which contradicts that $x$ and $y$ are independent.~\QED
 
\medskip

\begin{proposition}
\label{p:dimXOR}
If $\mathrm{dim}(x) = \sigma$ and $(x,y)$ are finitary-independent, then $\mathrm{dim}(x ~\mathrm{XOR}~ y) \geq \sigma$.
\end{proposition}
{\em Proof}.
 Note that $C(x \rest n \mid y \rest n) \leq C((x ~\mathrm{XOR}~ y) \rest n) + O(1)$, for all $n$ (this holds for all sequences $x$ and $y$).
Suppose there exists $\epsilon > 0$ such that $\mathrm{dim}(x ~\mathrm{XOR}~ y) \leq \sigma - \epsilon$.  It follows that, for infinitely many $n$, $C((x ~\mathrm{XOR}~ y) \rest n) \leq (\sigma - \epsilon) n$.
Then
\[
\begin{array}{ll}
C(x \rest n \mid y \rest n) & < C((x ~\mathrm{XOR}~ y) \rest n ) + O(1) \\ 
 & < (\sigma-\epsilon) n + O(1) \quad\quad \mbox{for infinitely many $n$}.
 \end{array}
 \]
 By the finitary-independence of $(x,y)$, $C(x \rest n) \leq C(x \rest n \mid y \rest n) + O(\log n) \leq (\sigma - \epsilon/2)n + O(1)$, $\mbox{i.o.}$ $n$, which contradicts the fact that $\mathrm{dim}(x) = \sigma$.~\QED
  
\medskip

 \begin{proposition}
 \label{p:weakindepXOR}
 \begin{itemize}
 \item[(a)] If $x$ is random and $(x,y)$ are finitary-independent, then $(y, x ~\mathrm{XOR}~ y)$ are finitary-independent.
 \item[(b)] If $x$ is random and $(x,y)$ are independent, then $(y, x ~\mathrm{XOR}~ y)$ are independent.
 \end{itemize}
 \end{proposition}
 {\em Proof}.
 We prove (a) ((b) is similar). Suppose that $y$ and $x ~{\mathrm{XOR}}~ y$ are not finitary-independent.
Then for every  
 constant $c$, there are infinitely many $n$, such that $C((x ~{\mathrm{XOR}}~ y)\rest n \mid y\rest n) < C((x ~{\mathrm{XOR}}~ y)\rest n) - c\log n$. Note that if a program   can produce $(x ~{\mathrm{XOR}}~ y)\rest n$ given $y\rest n$, then by doing an extra bitwise XOR with $y\rest n$ it will produce $x\rest n$. Thus,
$C(x\rest n \mid y\rest n) < C((x ~{\mathrm{XOR}}~ y)\rest n \mid y\rest n) + O(1)$ for all $n$. Combining with the first inequality,  for every  
 constant $c$ and for infinitely many $n$ we have:
\[
\begin{array}{ll}
C(x\rest n \mid y\rest n) & < C((x ~{\mathrm{XOR}}~ y)\rest n) - c\log n + O(1) \\
& < n - c \log n + O(1) \\
& < C(x\rest n) + 2 \log n - c\log n +O(1) \\
& = C(x \rest n) - (c-2) \log n +O(1).
\end{array}
\]
 This contradicts the fact that $x$ and $y$ are finitary-independent.~\QED
  
\medskip

\begin{proposition}

There are sequences $x, y$, and $z$ such that $(x,y)$ are independent,   $(x,z)$ are independent, but  $(x,y \oplus z)$ are not finitary-independent.

\end{proposition}
{\em Proof}.
  Take $y$ and $z$ two sequences that are random relative to each other, and let $x = y ~\mathrm{XOR}~ z$.
Then $(x,y)$ are independent, and $(x,z)$ are independent, by Proposition~\ref{p:weakindepXOR}. On the other hand note that $\mathrm{dim}(y ~\mathrm{XOR}~ z) = 1$ (by Proposition~\ref{p:dimXOR}) and  $C((y ~\mathrm{XOR}~ z) \rest n \mid (y \oplus z) \rest 2n) < O(1)$. Consequently, for every  
 constant $c$ and for almost every $n$, $C((y ~\mathrm{XOR}~ z) \rest n \mid (y \oplus z)\rest 2n) < C((y ~\mathrm{XOR}~ z) \rest n) - c(\log n + \log 2n)$, and thus, $(y ~\mathrm{XOR}~ z, y \oplus z)$ are not finitary-independent.~\QED

\medskip

In Remark~\ref{r:trivial}, we have listed several types of sequences that are independent or finitary-independent with any other sequence. The next result goes in the opposite direction: it exhibits a pair of sequences that can not be finitary-independent (and thus not independent).

\begin{proposition}{\rm \cite{frank}} 
\label{p:ce}
If $x$ and $y$ are left $\mbox{c.e.}$ sequences, ${\rm dim}(x) > 0$, and ${\rm dim}(y) > 0$, then $x$ and $y$ are not finitary-independent.
\end{proposition}
{\em Proof}.
 For each $n$, let ${\rm cm}_x(n) = \min\{ s \mid x_s \rest n = x \rest n \}$ and ${\rm cm}_y(n) = \min \{ s \mid y_s \rest n = y \rest n \}$ (the convergence moduli of $x$ and, respectively, $y$). Without loss of generality we can assume that ${\rm cm}_x(n) > {\rm cm}_y(n)$, for infinitely many $n$. For each $n$ satisfying the inequality, $y \rest n$ can be computed from $x \rest n$ as follows. First compute $s = {\rm cm}_x(n)$ (which can be done because $x \rest n$ is known) and output $y_s \rest n$. Consequently, for infinitely many $n$, $C( y \rest n \mid x \rest n) < O(1)$. On the other hand, since $\dim(y)>0$, there exists a constant $c$ such that $C(y \rest n) \geq c \cdot n$, for almost every $n$. Consequently, $x$ and $y$ are not finitary-independent.~\QED

\section{Examples of independent and finitary-independent sequences} 
We give examples of pairs of sequences that are independent or finitary-independent (other than the trivial examples from Remark~\ref{r:trivial}).

 \begin{theorem} 
 \label{t:indepex}
 Let $x$ be a random sequence and let $y$ be a sequence that is random relative to $x$. Then $x$ and $y$ are independent.
 \end{theorem}
 {\em Proof}.
 Since $y$ is random relative to $x$, for all $n$, 
 $C^x(y \rest n) > n - 2 \log n - O(1) \geq C(y \rest n) - 2 \log n  - O(1)$.
 The van Lambalgen Theorem \cite{vlam:j:randomness} implies that $x$ is random relative to $y$ as well. Therefore, in the same way, $C^y(x \rest n)  > n - 2 \log n  - O(1) \geq C(x \rest n) - O(\log n)$.~\QED

\medskip

 From Theorem~\ref{t:indepex} we can easily derive examples of pairs $(x,y)$ that are independent and which have constructive Hausdorff dimension $\epsilon$, for every  
 rational $\epsilon > 0$. For example, if we start with $x$ and $y$ that are random with respect to each other and build $x' = x(1)~0 x(2)~0 \ldots$ ($\mbox{i.e.,}$ we insert $0$s in the even positions) and similarly build $y'$ from $y$, then $x'$ and $y'$ have constructive Hausdorff dimension equal to $1/2$ and are independent (because $C^{x'}(y' \rest n)$ and $C^{x}(y \rest (n/2))$ are within a constant of each other, as are $C(y' \rest n)$ and $C(y \rest (n/2))$).
 The pairs of sequences from Theorem~\ref{t:indepex} (plus those derived from there as above) and those from Remark~\ref{r:trivial} are the only examples of independent sequences that we know. Thus, currently, we have examples of independent pairs $(x,y)$ only for the case when $x$ has maximal prefix-free complexity ($\mbox{i.e.}$, $x$ is random) or $x$ is obtained via a straightforward transformation as above from a random sequence, and for the case when $x$ has minimal prefix-free complexity ($\mbox{i.e.}$, $x$ is $H$-trivial). We believe that for every $x$, there are sequences $y$ independent with it, and moreover we believe that the set of sequences independent with $x$ has measure one. For finitary-independence these facts are true.

\begin{theorem} Let $x$ be an arbitrary sequence and let $y$ be a sequence that is random conditioned by $x$. Then $x$ and $y$ are finitary-independent.
\end{theorem}

{\em Proof}.
 Suppose $x$ and $y$ are not finitary-independent. Then there are infinitely many $n$ with
$C(y\rest n \mid x\rest n) < C(y\rest n) - 5 \log n$.
Consider a constant $c_1$ satisfying $C(y\rest n) < n+ c_1$, for all $n$. We get (under our assumption) that, for infinitely many $n$.
$C(y\rest n \mid x\rest n) < n  - 5 \log n + c_1$.
Then, by inequality~\ref{e:lifting}, for infinitely many $n$,
$C^{x\rest n}(y\rest n) < n - 3 \log n + c + c_1$.
Note that that for every $n$ and every $m \geq n$, $
C^{x\rest m}(y\rest n ) < C^{x\rest n}(y\rest n )$.
Thus, for infinitely many $n$ and for all $m > n$,
\begin{equation}
\label{e:ineq1}
C^{x\rest m}(y\rest n ) < n  - 3 \log n + (c + c_1).
\end{equation}
On the other hand, $y$ is random conditioned by $x$. Therefore, for all $n$, $H^x (y\rest n)> n - O(1)$. Let $U'$ be the universal machine underlying the complexity $H(\cdot)$ and let $p^*$ be the shortest program such that $U'^x (p^*) = y\rest n$ (if there are ties, take $p^*$ to be the lexicographically smallest among the tying programs). Let $m(n) = \min (n, \mbox{use} (U'^x(p*)))$. Note that, for all $n$, $H^x (y\rest n) = H^{x\rest m(n)} (y\rest n)$. It follows that, for every $n$,
$H^{x\rest m(n)}(y\rest n )=  H^x (y\rest n)> n - O(1)$.
Recall that for every strings $u$ and $v$, $C^{v}(u) > H^{v}(u ) - 2 \log |u| - O(1)$. Thus, for every $n$,
\begin{equation}
\label{e:ineq2}
C^{x\rest m(n)}(y\rest n ) > n - 2 \log n - O(1).
\end{equation}
Inequalities~(\ref{e:ineq1}) and~(\ref{e:ineq2}) are contradictory.~\QED

\begin{theorem}
\label{t:measone}
For every  
 $x$, the set $\{y \mid y \mbox{ finitary-independent with } x \}$ has measure one.
\end{theorem}

{\em Proof}.
 By the previous result, the  set in the statement of the theorem contains the set $\{y \mid y \mbox{ random conditioned by } x \}$ which has measure one.~\QED

\medskip

Thus there are many (in the measure-theoretical sense) pairs of sequences that are finitary-independent.  But is it possible to have such pairs satisfying a given constraint? We consider one instance of this general issue.

\begin{proposition}
\label{p:constraint}
If $x$ is a random sequence then there are $y$ and $z$ such that $(y,z)$ are finitary-independent and $x = y~\mathrm{XOR}~z$.
\end{proposition}
{\em Proof}.  Take a sequence $y$ finitary-independent with $x$.
Then, by Proposition~\ref{p:weakindepXOR}, $y$ and $(x ~{\mathrm{XOR}}~ y)$ are finitary-independent. By taking $z = x ~{\mathrm{XOR}}~ y$, it follows that $x = y ~{\mathrm{XOR}}~ z$, with $y$ and $z$ finitary-independent.~\QED

\section{Effective constructions of finitary-independent sequences}

The examples of (finitary-) independent sequences that we have provided so far are  existential (\ie, non-constructive). In this section we investigate to what extent it is possible to effectively construct such sequences. We show some impossibility results and therefore we focus on the weaker type of independence, finitary-independence (clearly, if it is not possible to produce a  pair of sequences  that are finitary-independent, then it is also not possible to produce a pair of sequences that are independent). Since a C-logarithmic sequence is finitary-independent with any other sequence, the issue of constructibility is interesting if we also require that the sequences have complexity above that of C-logarithmic sequences (see Remark~\ref{r:trivial}). Such sequences are of course non-computable, and therefore the whole issue of constructibility appears to be a moot point. However this is not so if we assume that we already have in hand one (or several) non-computable sequence(s), and we want to build additional sequences that are finitary-independent. Informally speaking, we investigate the following questions:
\smallskip

{\bf Question (a)} Is it possible to effectively construct from a sequence $x$ another sequence $y$ $\mbox{(finitary-)}$ ~independent with $x$, where the independence is not trivial (recall Remark~\ref{r:trivial})? This question has two variants depending on whether we seek a uniform procedure (\ie, one procedure that works for all $x$), or whether we allow the procedure to depend on $x$. 
\smallskip

{\bf Question (b)} Is it possible to effectively construct from a ~sequence $x$ two sequences $y$ and  $z$ that are $\mbox{(finitary-)}$~independent, where the independence is not trivial? Again, there are uniform and non-uniform variants of this question.
\smallskip

We analyze these questions in Section~\ref{s:onesource}.
Similar questions for the case when the input consists of two sequences $x_1$ and $x_2$ are tackled in Section~\ref{s:twosources}.

\subsection{If we have one source}
\label{s:onesource}
We first consider the uniform variant of Question~(a): Is there a Turing reduction $f$ such that for all $x \in \{0, 1\}^*$, $(x, f(x))$ are (finitary-)~independent? We even relax the requirement and demand that $f$ should achieve this objective only if $x$ has positive constructive Hausdorff dimension (this only makes the following impossibility results stronger).

As discussed above, we first eliminate some trivial instances of this question.
Without any requirement on the algorithmic complexity of the desired $f(x)$, the answer is trivially YES because we can take $f(x) = 0^\omega$ (or any other computable sequence).
Even if we only require that $f(x)$ is not computable, then the answer is still trivially YES because we can make $f(x)$ to be C-logarithmic. For example, consider $$f(x) =x(1) ~x(2)0 ~x(3)000 \ldots ~x(k) \underbrace{0 \ldots 0}_{2^{k-1}-1} \ldots .$$ Then $f(x)$ is C-logarithmic, but not computable provided $x$ is not computable, and $(x, f(x))$ are finitary-independent simply because $f(x)$ is C-logarithmic.

As noted above, the question is interesting if we require $f(x)$ to have some ``significant" amount of randomness whenever $x$ has some ``significant" amount of randomness. We expect that in this case the answer should be negative, because, intuitively, one should not be able to produce independence (this is property P3 in Remark~\ref{r:properties}). 

We consider two situations depending on two different meanings of the concept of ``significant" amount of randomness.

\medskip

{\bf Case 1:} We require that $f(x)$ is not C-logarithmic. We do not solve the question, but we show that every  
reduction $f$ that potentially does the job must have non-polynomial use.

\begin{proposition}
\label{p:notindep}
Let $f$ be a Turing reduction. For every  
 sequence $x$, if the function $use_f^x(n)$ is polynomially bounded, then  $x$ and $f(x)$ are not finitary-independent, unless one of them is C-logarithmic.
\end{proposition}

{\em Proof}.
 Let $y$ be $f(x)$. Then for every $n$, let $m(n) = \max_{k \leq n} use_f^x(1^n))$. Then $y\rest n$ depends only on $x\rest m(n)$ and $m(n)$ is polynomial in $n$. Then $C(y\rest n \mid x\rest m(n)) \leq O(\log n)$. If $x$ and $y$ were finitary-independent, then $C(y\rest n) \leq C(y \rest(n) \mid x \rest m(n)) + O(\log n + \log m(n)) \leq O(\log(n)) + \log(m(n)) \leq O(\log n)$, for all $n$, \ie,  $y$ would be C-logarithmic
.~\QED
 
\medskip

{\bf Case 2:} We require that $f(x)$ has complexity just above that of C-logarithmic sequences (in the sense below).   We show that in this case, the answer to the uniform variant of Question (a) is negative: there is no such $f$.   The following definition introduces a class of sequences having complexity just above that of C-logarithmic sequences.
\begin{definition}
\label{d:simpleio}
A sequence $x$ is C-superlogarithmic if for every constant $c > 0$, $C(x \rest n) > c \log n$, for almost every $n$.
\end{definition}
The next proofs use the following facts.
\begin{fact}
{\rm (Variant of Theorem 3.1 in~\cite{nie-rei:c:wtt-Kolm-increase})}
\label{f:NiesReimann}
For all rationals $0 \leq \alpha < \beta < 1$, and for every  set $S$ that is infinite and computable, there exists a sequence $x$ such that $\dim(x) = \alpha$ and for all wtt-reductions $f$, either $f(x)$ does not exist or $C(f(x)\rest n) \leq \beta n$, for infinitely many $n$ in $S$.
\end{fact}
\begin{fact}
\label{f:bds}
{\rm (Variant of Theorem 3.1 in~\cite{bie-dot-ste:c:haussdimension})} For every Turing reduction $h$, for all rationals $0 < \alpha < \beta < 1$,  and for every  set $S$ that is infinite and computable, there is a sequence $x$ with  $\dim(x) \geq \alpha$ such that either $h(x)$ does not exist or $C(h(x)\rest n) < \beta n$, for infinitely many $n$ in $S$.
\end{fact}
\begin{fact}
\label{f:one-two-sources}
{\rm (Theorem 4.15 in~(\cite{zim:t:extractKolm})} 
For any $\delta > 0$, there exist a constant $c$, a set $S$ that is infinite and computable,  and a truth-table reduction $g: \{0,1\}^\infty \times \{0,1\}^\infty \rightarrow \{0,1\}^\infty$ (\ie, $g$ is a Turing machine with two oracles) with the following property: 

If the input sequences $x$ and $y$  are finitary-independent and satisfy $C(x\rest n) > c \cdot \log n$ and $C(y\rest n) > c \cdot \log n$, for almost every $n$, then the output $z = f(x,y)$ satisfies $C(f(x,y) \rest n) > (1- \delta) \cdot n$, for almost every $n$ in $S$.
\end{fact}
Theorem 3.1 in~\cite{nie-rei:c:wtt-Kolm-increase} is for $S = {\mathbb N}$ (and is stronger in that $\alpha = \beta$) but its proof can be modified in a straightforward manner to yield Fact~\ref{f:NiesReimann}. Theorem 3.1 in~\cite{bie-dot-ste:c:haussdimension} is also for $S = {\mathbb N}$ and can also be modified in a simple manner -- using Fact~\ref{f:NiesReimann} -- to yield Fact~\ref{f:bds}.

We can now state the impossibility results related to  {\bf Case 2}. To simplify the structure of quantifiers in the statement of the following result, we posit here the following task for a function $f$ mapping sequences to sequences:
 
\smallskip

TASK A: for every $x \in \{0,1\}^\infty$ with ${\rm dim}(x) > 0$,  the following should hold: 
\begin{itemize}
  \item[(a)] $f(x)$ exists.
	\item[(b)] $f(x)$ is C-superlogarithmic.
	\item[(c)] $x$ and $f(x)$ are finitary-independent.
\end{itemize}

\begin{theorem} 
\label{t:impossible-one-source}
There is no Turing reduction $f$ that satisfies TASK A.
\end{theorem}

{\em Proof}.
Suppose there exists $f$ satisfying (a), (b) and (c) in TASK A.  Let $S$ be the infinite, computable set and let $g$ be the truth-table reduction promised by Fact~\ref{f:one-two-sources} for $\delta = 0.3$. Let $h$ be the Turing reduction $h(x) = g(x, f(x))$. Let $x^*$ be the sequence promised by Fact~\ref{f:bds} for $\alpha = 0.5$, $\beta = 0.6$, and the above set $S$ and Turing reduction $h$.  On one hand, by Fact~\ref{f:bds}, $C(h(x^*) \rest n) < 0.6n$, for infinitely many $n \in S$.  On the other hand, by Fact~\ref{f:one-two-sources}, $C(h(x^*) \rest n) > 0.7n$, for almost every $n \in S$. We have reached a contradiction.~\QED
 
\medskip

We next consider the uniform variant of Question~(b).
\smallskip

First we remark, that by van Lambalgen Theorem \cite{vlam:j:randomness}, if the sequence $x$ is random, then $x_{even}$ and $x_{odd}$ are random relative to each other  (where $x_{odd}$ is $x(1) x(3) x(5) \ldots $ and $x_{even}$ is $x(2) x(4) x(6) \ldots$). Thus, $x_{even}$ and $x_{odd}$ are certainly independent.

Kautz~\cite{kau:t:kolmog} has shown a much more general result by examining the splittings of sequences obtained using bounded Kolmogorov-Loveland selection rules.\footnote{A Kolmogorov-Loveland selection rule is an effective process for selecting bits from a sequence. Informally, it is an iterative process and at each step, based on the bits that have been already read, a new bit from the sequence is chosen to be read and (before that bit is actually read) the decision on whether that bit is selected or not is taken. A \emph{bounded} Kolmogorov-Loveland selection rule satisfies a certain requirement of monotonocity for deciding the selected bits, see~\cite{kau:t:kolmog}.}  He showed that if $x$ is a random sequence, $x_0$ is the subsequence of $x$ obtained by concatenating the bits of $x$ chosen by an arbitrary bounded Kolmogorov-Loveland selection rule, and $x_1$ consists of the bits of $x$ that were not selected by the selection rule, then $x_0$ and $x_1$ are random with respect to each other (and thus independent).

We show that the similar result for sequences with constructive Hausdorff dimension $\sigma \in (0,1)$ is not valid. In fact, our next result is stronger, and essentially gives a negative answer to the uniform variant of Question~(b).

We posit the following task for two functions $f_1$ and $f_2$ mapping sequences to sequences:

\smallskip

TASK B: for every $x \in \{0,1\}^\infty$ with ${\rm dim}(x) > 0$,  the following should hold:
\begin{itemize}
 \item[(a)] $f_1(x)$ and $f_2(x)$ exist,
	\item[(b)] $f_1(x)$ and $f_2(x)$ are C-superlogarithmic, 
	\item[(c)] $f_1(x)$ and $f_2(x)$ are finitary-independent.
\end{itemize}

\begin{theorem} 
\label{t:impossible-two-sources}
There are no Turing reductions $f_1$ and $f_2$ satisfying TASK B.
\end{theorem}
{\em Proof}.  Similar to the proof of Theorem~\ref{t:impossible-one-source}.~\QED
 
\medskip

The non-uniform variants of Questions ~(a) and~(b) remain open. In the particular case when $f$ is a wtt-reduction, we present  impossibility results analogous to those in Theorem~\ref{t:impossible-one-source} and Theorem~\ref{t:impossible-two-sources}. The proofs are similar, with the difference that we use Fact~\ref{f:NiesReimann} instead of Fact~\ref{f:bds}.

\begin{theorem}
\label{t:wtt-1}
For all rational $\sigma \in (0,1)$, there exists $\dim (x) = \sigma$ such that for every  
 wtt-reduction $f$, at least one of the following holds true:
\begin{itemize}
\item[(a)] $f(x)$ does not exist,
\item[(b)] $f(x)$ is not finitary-independent with $x$,
\item[(c)] $f(x)$ is not C-superlogarithmic.
\end{itemize}
\end{theorem}

\begin{theorem}
\label{t:unif-impossible-two-sources}
For all rational $\sigma \in (0,1)$, there exists $x$ with $\dim (x) = \sigma$ such that for every  
 wtt-reductions $f_1$ and $f_2$, at least one of the following holds true:
\begin{itemize}
\item[(a)] $f_1(x)$ does not exist or $f_2(x)$ does not exist,
\item[(b)] $f_1(x)$ and $f_2(x)$ are not finitary-independent,
\item[(c)] $f_1(x)$ is not C-superlogarithmic or $f_2(x)$ is not C-superlogarithmic.
\end{itemize}
\end{theorem}
Theorem~\ref{t:unif-impossible-two-sources} has an interesting implication regarding sequences with constructive Hausdorff dimension in the interval $(0,1)$. Suppose, for example that we want to construct a sequence with constructive Hausdorff dimension 1/2. The first idea that comes to mind is to take a random sequence $x = x(1) x(2) \ldots$ and either  consider the sequence $y = x(1) 0 x(2) 0  \ldots$ (we insert $0$s in all even positions) or the sequence $z= x(1)x(1) x(2) x(2) \ldots$ (we double every bit). The sequences $y$ and $z$ have constructive Hausdorff dimension 1/2. 
Theorem~\ref{t:unif-impossible-two-sources} shows, roughly speaking, that there are sequences with dimension strictly between $0$ and $1$, where the partial randomness is due necessarily to one of the two methods stated above. Formally, for every  
 rational $\sigma \in (0,1)$, there is a sequence $x$ with $\mathrm{dim}(x) = \sigma$ so that no matter what wtt method we use for selecting from $x$ two subsequences,  either one of the resulting subsequences has low  complexity or the two resulting subsequences are not independent.

\subsection{If we have two sources}
\label{s:twosources}

We have seen some limits on the possibility of constructing a finitary-independent sequences starting from one sequence. What if we are given two finitary-independent sequences: is it possible to construct from them more finitary-independent sequences?

First we observe that if $x$ and $y$ are two independent sequences and $g$ is an arbitrary Turing reduction, then it does not necessarily follow that $x$ and $g(y)$ are independent (as one may expect). On the other hand it does follow that $x$ and $g(y)$ are finitary-independent.

\begin{proposition}{\rm \cite{frank}}\label{p:createdep}
There are two independent sequences $x$ and $y$ and a Turing reduction $g$ such that $x$ and $g(y)$ are not independent.
\end{proposition}
{\em Proof}.
 Let $z$ be a random sequence and let $u, v$, and $w$ be sequences such that $z = u \oplus v \oplus w$. By van Lambalgen Theorem \cite{vlam:j:randomness}, each of the sequences $u, v$, and $w$ are random relative to the join of the other two. We define the sequences $x$ and $y$ as follows:
\[
\begin{array}{ll}
x(2^n) & = u(n), \mbox{ for all $n \in \nat$} \\
x(m) &= v(m), \mbox{ for every $m$ that is not a power of $2$}\\
y(2^n) & = u(n), \mbox{ for all $n \in \nat$} \\
y(m) &= w(m), \mbox{ for every $m$ that is not a power of $2$}
\end{array}
\]
\begin{claim} The sequences  $x$ and $y$ are independent.
\end{claim}
{\em Proof}.
 Suppose $x$ and $y$ are not independent. Then, similarly to Proposition~\ref{p:cond} (e), for infinitely many $n$, 
$C^x (y \rest n) < C(y \rest n) - 7 \log n$.
Then
\[
\begin{array}{ll}
C^{u \oplus v}(w \rest n) & \leq C^{u \oplus v}(y \rest n) + 2 \log n + O(1)  \\
												& \quad\quad				\mbox{(because $w\rest n$ and $y \rest n$ differ in only $\log n$ bits)}\\

									&	\leq C^x  (y \rest n) + 2 \log n + O(1)   \\
														& \quad \quad				\mbox{(because queries to $x$ can be replaced 																																by queries to $u$ and $v$)} \\
										
									&	\leq C(y \rest n) - 7 \log n + 2 \log n +O(1), \\
									& \quad \quad									\mbox{for infinitely many $n$} \\
										
									&	\leq C(w \rest n) + 2 \log n - 7 \log n + 2 \log n +O(1) \\

								&		= C(w \rest n) - 2 \log n +O(1) \\
										
								&		\leq n - 3 \log n + O(1).
\end{array}
\]
										
This contradicts that w is random with respect to $u \oplus v$.~\QED
 
\medskip

It is easy to define a Turing reduction $g$ such that $g(y) = u$. Notice that $C^x(u \rest n) = O(\log n)$, because $u$ is many-one reducible to $x$. On the other hand $C(u \rest n) \geq n - 2\log n +O(1)$, for every $n$, because $u$ is random. Therefore $x$ and $g(y)$ are not independent.~\QED

We do not know if the facts that $x$ and $y$ are finitary-independent and $g$ is a Turing reduction, imply that $x$ and $g(y)$ are finitary-independent  This would show that finitary-dependency cannot be created. 

The following weaker result holds.

\begin{proposition}
\label{p:no-dependency}
If $x$ and $y$ are independent, and $g$ is a Turing reduction, then $x$ and $g(y)$ are finitary-independent (provided $g(y)$ exists).
\end{proposition}
{\em Proof}.
 Since $x$ and $y$ are independent, there exists a constant $c$ such that for all n,
\[
C^y(x \rest n) \geq C(x \rest n) - c \log n.
\] 
Suppose that $x$ and $g(y)$ are not finitary-independent. Then there are infinitely many $n$ such that
$C(x\rest n \mid g(y)\rest n) < C(x \rest n) - (c+4) \log n$.
Since $C^y(x \rest n) \leq C(x \rest n \mid g(y)\rest n) + 2 \log n + O(1)$, it would follow that, for infinitely many $n$,
\[
C^y(x \rest n) \leq C(x \rest n) -  (c+1)\log n,
\]
which contradicts the first inequality.~\QED
 
\medskip

\begin{corollary}
\label{c:windepnotindep}
There are sequences that are finitary-independent but not independent.
\end{corollary}
{\em Proof}.
 The sequences $x$ and $g(y)$ from Proposition~\ref{p:createdep} are not independent, but they are finitary-independent by Proposition~\ref{p:no-dependency}.~\QED
 
\medskip

As we mentioned, we do not know if Proposition~\ref{p:no-dependency} can be strengthened to hold if $x$ and $y$ are finitary-independent. However, for such $x$ and $y$, there exists a simple procedure that starting with the pair $(x,y)$, produces a new pair of finitary-independent sequences. Namely, we take the pair $(x, y_{odd})$.

\begin{proposition} If $x$ and $y$ are finitary-independent, then $x$ and $y_{odd}$ are finitary-independent.
\end{proposition}

{\em Proof}.
 Suppose that for every  
 constant $c$ there are infinitely many $n$ such that
$C(x \rest n \mid y_{odd} \rest n) < C(x \rest n) - c \cdot \log n$. Note that, for all $n$, 
$C(x \rest n \mid y \rest 2n) \leq   C(x \rest n \mid y_{odd} \rest n) +O(1)$.  Our assumption implies that for every  
 constant $c$ there are infinitely many $n$ such that $C(x \rest n \mid y \rest 2n) < C(x \rest n) - c \log n +O(1)$. By Proposition~\ref{p:cond}, (a), this contradicts the fact that $x$ and $y$ are finitary-independent.~\QED
 
\medskip

The next issue that we study is whether given a pair of $\mbox{(finitary-)independent}$ strings $(x,y)$, it is possible to effectively produce another string that is $\mbox{(finitary-)independent}$ with both $x$ and $y$. We give a positive answer for the case when  $x$ and $y$ are both  random. The similar question for non-random $x$ and $y$ remains open (but see Section~\ref{s:finiteindep}).

\begin{theorem}
\label{t:doubleindep}
There exists an effective transformation $f$ with polynomially-bounded use such that if $x$ and $y$ are random and independent (respectively finitary-independent), then $f(x,y)$ is independent (respectively, finitary-independent) with both $x$ and $y$, and the independence is not trivial (recall Remark~\ref{r:trivial}). 
\end{theorem}

{\bf Remark:} Contrast with Proposition~\ref{p:notindep}, where we have shown that for every  
 $x$, for every  
 effective transformation $f$ with polynomially-bounded use, $x$ and $f(x)$ are not finitary-independent.
\smallskip

{\em Proof}.
 We take $f(x,y) = x ~{\mathrm{XOR}}~ y$ and take into account Proposition~\ref{p:weakindepXOR}.~\QED

\subsection{Producing independence: the finite case}
\label{s:finiteindep}
An interesting issue is whether given as input several sequences that are (finitary-) independent, there is an effective way to construct a sequence that is (finitary-) independent with each sequence in the input (and the independence is not trivial). A result of this type is obtained for the case when the input consists of two random sequences $x$ and $y$ in Theorem~\ref{t:doubleindep}.
We do not know if in Theorem~\ref{t:doubleindep} we can remove the assumption that $x$ and $y$ are random.  

In what follows we will  consider the simpler case of strings. In this setting we are able to give a positive answer for the situation when we start with three\footnote{The case when the input consists of two independent strings remains open.} input strings that are independent (and not necessarily random). First we define the analogue of independence for  strings.

\begin{definition}
Let $c \in \real^+$ and $k \in \nat$. We say that strings $x_1, x_2, \ldots, x_k$ in $\{0,1\}^*$ are $c$-independent if
\[
C(x_1 x_2 \ldots x_k) \geq C(x_1) + C(x_2) + \ldots + C(x_k) - c (\log|x_1| + \log|x_2| + \ldots + \log|x_k|).
\]
\end{definition}

\if01
{\bf  $C(x_k) - c (\log|x_1|x_2| \cdots |x_k|)$ is not better than $C(x_k) - c (\log|x_1| + \log|x_2| + \ldots + \log|x_k|)$??}
\fi

The main result of this section is the following theorem, whose proof draws from the techniques of~\cite{zim:t:extractKolm}.

\begin{theorem}
\label{t:stringindep}
For all constants $\sigma > 0$ and $\sigma_1 \in (0, \sigma)$, there exists a computable function $f: \{0,1\}^* \times \{0,1\}^* \times \{0,1\}^* \rightarrow \{0,1\}^*$ with the following property:
For  every  
 $c \in \real^+$ there exists $c \in \real^+$ such that if the input consists of a triplet of $c$-independent strings having sufficiently large length $n$ and  plain complexity at least $\sigma \cdot n$, then the output is $c$-independent with each element in the input triplet and has length $\lfloor \sigma_1 n \rfloor$.

More precisely, if 
\begin{itemize}
\item[\rm (i)]$(x,y,z)$ are $c$-independent, 
\item[\rm (ii)]
$|x| = |y| = |z| = n$, and 
\item[\rm (iii)]$C(x) \geq \sigma \cdot n$, $C(y) \geq \sigma \cdot n$, $C(z) \geq \sigma \cdot n$,
\end{itemize} 
then, provided $n$ is large enough, the following pairs of strings
$(f(x,y,z), x)$,  $(f(x,y,z), y)$, $(f(x,y,z), z)$ are $c$-independent, $|f(x,y,z)| = \lfloor \sigma_1 n\rfloor$, and $C(f(x,y,z)) \geq \lfloor \sigma_1 n\rfloor - O(\log n)$.
\end{theorem}
Before we delve into the proof, we establish several preliminary facts.

\begin{lemma}
\label{l:threeindep}
If $x_1, x_2, x_3$ are three strings that are $c$-independent, then
\[
C(x_1 \mid x_2 x_3) \geq C(x_1) - (c+2)(\log|x_1| + \log |x_2| + \log |x_3|) - O(1).
\]
\end{lemma}

{\em Proof}.
The following inequalities hold for  every  
 three strings and in particular for the strings $x_1$, $x_2$, and $x_3$:
\[
C(x_1 x_2 x_3) \leq C(x_2 x_3) + C(x_1 \mid x_2 x_3) + 2 \log |x_1| +O(1),
\]
and
\[
C(x_2 x_3) \leq C(x_2) + C(x_3) + 2 \log|x_2| + O(1).
\]
Then
\[
\begin{array}{ll}
C(x_1 \mid x_2 x_3) & \geq C(x_1 x_2 x_3) - C(x_2 x_3) - 2 \log |x_1| - O(1) \\ 
 & \geq C(x_1) + C(x_2) + C(x_3) - c (\log |x_1| + \log |x_2| + \log |x_3|) \\
 &\phantom{x} - (C(x_2) + C(x_3) + 2 \log |x_2| + O(1)) - 2 \log |x_1| - O(1) \\ 
 & \geq C(x_1) - (c+2) (\log|x_1| + \log |x_2| + \log |x_3|) - O(1).
 \end{array}
 \]
 ~\QED
  
\medskip

 The next lemma establishes a combinatorial fact about the possibility of coloring the cube
 $[N] \times [N] \times [N]$ with $M$ colors such that  every  
 planar rectangle contains all the colors in about the same proportion. Here $N$ and $M$ are natural numbers, $[N]$ denotes the set
 $\{1, 2, \ldots, N\}$,  $[M]$ denotes the set
 $\{1, 2, \ldots, M\}$and a planar rectangle is a subset of  $[N] \times [N] \times [N]$ having one of the following three forms: $B_1 \times B_2 \times \{k\}$, $B_1 \times \{k\} \times B_2$, or $\{k\} \times B_1  \times B_2$, where $k \in [N]$, $B_1 \subseteq [N]$ and $B_2 \subseteq [N]$. 
 
 \begin{lemma}
 \label{l:coloring}
 Let $0 < \sigma_1 < \sigma_2 < 1$. For  every  
 $n$ sufficiently large, it is possible to color the cube $[2^n] \times [2^n] \times [2^n]$ with $M = 2^{\lfloor \sigma_1 n \rfloor}$ colors in such a way that  every  
 planar rectangle satisfying $\lll B_1 \rrr = a 2^{\lceil \sigma_2 n \rceil}$ and
 $\lll B_2 \rrr = b 2^{\lceil \sigma_2 n \rceil}$ for some natural numbers $a$ and $b$ contains at most $(2/M) \lll B_1 \rrr \lll B_2 \rrr$ occurrences of color $c$, for  every  
 color $c \in [M]$.
 \end{lemma}
 
 {\em Proof}.
 We use the probabilistic method. Let $N = 2^n$. We color each cell of the $[N] \times [N] \times [N]$  cube with one color chosen independently and uniformly at random  from $[M]$. For $i,j,k \in [N]$, let $T(i,j,k)$ be the random variable that designates the color of the cell $(i,j,k)$ in the cube. For  every  
 fixed cell $(i,j,k)$ and for  every  
 fixed color $c \in [M]$, ${\rm Prob}(T(i,j,k) = c) = 1/M$, because the colors are assigned independently and uniformly at random. Let us first consider some fixed subsets $B_1$ and $B_2$ of $[N]$ having size $2^{\lceil \sigma_2 n \rceil}$, a fixed $k \in [N]$, and a fixed color $c \in [M]$. Let $A$ be the event ``the fraction of occurences of $c$ in the planar rectangle $B_1 \times B_2 \times \{k\}$ is greater than $2/M$." Using the Chernoff bounds, it follows that $${\rm Prob}(A) < e^{-(1/3) (1/M) N^{2\sigma_2}}.$$
 The same upper bounds hold for the probabilities of the similar events regarding the planar rectangles $B_1 \times \{k\} \times B_2$ and $\{k\} \times B_1 \times B_2$. Thus, if we consider the event $B$ ``there is some color with a fraction of appearances in one of the three planar rectangles mentioned above   greater than $(2/M)$", then, by the union bound, 
 \begin{equation}
 \label{e:union}
 {\rm Prob}(B) < 3M e^{-(1/3) (1/M) N^{2\sigma_2}}.
 \end{equation}
 The number of ways to choose $B_1 \subseteq [N]$ with $\lll B_1 \rrr = 2^{\lceil \sigma_2 n \rceil}$, 
 $B_2 \subseteq [N]$ with $\lll B_2 \rrr = 2^{\lceil \sigma_2 n \rceil}$ and $k \in [N]$ is approximately (ignoring the truncation) ${ N \choose N^{\sigma_2}} \cdot{ N \choose N^{\sigma_2}} \cdot N$, which is bounded by 
 \begin{equation}
 \label{e:comb}
 e^{2N^{\sigma_2}} \cdot e^{2N^{\sigma_2} (1-\sigma_2) \ln(N)} \cdot e^{\ln N},
 \end{equation} (we have used the inequality ${n \choose k} < (en/k)^k$). Clearly, for our choice of $M$, the right hand side in~(\ref{e:comb}) times the right hand side in~(\ref{e:union}) is less than $1$. It means that there exists a coloring where no color appears a fraction larger than $(2/M)$ in  every  
 planar rectangle with $B_1$ and $B_2$ having size exactly
 $2^{\lceil \sigma_2 n \rceil}$. For planar rectangles having the sizes of $B_1$ and $B_2$ an integer multiple of $2^{\lceil \sigma_2 n \rceil}$, the assertion holds as well because such rectangles can be partitioned into subrectangles having the size exactly $2^{\lceil \sigma_2 n \rceil}$.~\QED

\medskip

 {\em Proof} of {\bf Theorem~\ref{t:stringindep}.} We take $n$ sufficiently large so that all the following inequalities hold. Let $x^*$, $y^*$ and $z^*$ be a triplet of strings of length $n$ satisfying the assumptions in the statement. Let $N = 2^n$ and let us consider a constant $\sigma_2 \in (\sigma_1, \sigma)$. By exhaustive search we find a coloring $T:  [N] \times [N] \times [N] \rightarrow [M]$ satisfying the properties in Lemma~\ref{l:coloring}. 
 Identifying the strings $x^*$, $y^*$ and $z^*$ with their indeces in the lexicographical ordering of $\Sigma^n$, we define $w^* = T(x^*, y^*, z^*)$. Note that the length of $w^*$ is $\log M = \lfloor \sigma_1 n \rfloor$, which we denote $m$. We will show that $C(w^* \mid z^*) \geq m - c' \log m$, for $c' = 3c + d + 13$, for a constant $d$ that will be specified later. Since $C(w^*) \leq m + O(1)$, it follows that $w^*$ and $z^*$ are independent. In a similar way, it can be shown that $w^*$ and $x^*$ are independent, and $w^*$ and $y^*$ are independent.
 
 For the sake of obtaining a contradiction, suppose that $C(w^* \mid z^*) < m - c' \log m$.
 The set $A = \{w \mid C(w \mid z^*) < m - c' \log m \}$ has size $< 2^{m - c' \log m}$  and, by our assumption, contains $w^*$. 
 
 Let $t_1$ be such that $C(x^*) = t_1$ and $t_2$ be such that $C(y^* \mid z^*) = t_2$.  Note that $t_1 > \sigma_2 n$. The integer $t_2$ is also larger than $\sigma_2 n$, because $C(y^* \mid z^*) \geq C(y^* \mid z^* x^*) - 2 \log n - O(1) \geq C(y^*) - (c+4) (3 \log n) - O(1) \geq \sigma n - (3c+12) \log n - O(1)> \sigma_2 n$. For the second inequality we have used Lemma~\ref{l:threeindep}.
 
 Let $B_1 = \{x \in \Sigma^n \mid C(x) \leq t_1\}$. Note that $B_1$ has size bounded by $2^{t_1 + 1}$. We take a set $B_1'$ including $B_1$ having size exactly $2^{t_1 + 1}$. Similarly, let
 $B_2 = \{y \in \Sigma^n \mid C(y \mid z^*) \leq t_2\}$ and let $B_2'$ be a set that includes $B_2$ and has size exactly $2^{t_2+1}$. Let $k$ be the index of $z^*$ in the lexicographical ordering of $\Sigma^n$. By Lemma~\ref{l:coloring}, it follows that for  every  
 $a \in [M]$,
 $$\lll T^{-1} (a) \cap (B_1' \times B_2' \times \{k\}) \rrr \leq (2/M) \lll B_1' \rrr \lll B_2' \rrr .$$ Consequently,
 \[
 \begin{array}{ll}
 \lll T^{-1}(A)  \cap (B_1 \times B_2 \times \{k\}) \rrr & \leq  \lll T^{-1}(A)  \cap (B_1' \times B_2' \times \{k\}) \rrr \\[1ex]
 & = \sum_{a \in A} \lll T^{-1}(a)  \cap (B_1' \times B_2' \times \{k\}) \rrr \\[1ex]
 & < 2^{m - c' \log m} \cdot (2/2^m) \lll B_1' \rrr \lll B_2' \rrr = 2^{t_1 + t_2 + 3 - c' \log m}.
 \end{array}
 \]
 Note that given $z^*$, $m - c' \log m$, $t_1$ and $t_2$, we can enumerate $T^{-1}(A)  \cap (B_1 \times B_2 \times \{k\})$. Since
 $(x^*, y^*, z^*)$ is in this set, it follows that the complexity of $x^*y^*$ given $z^*$ is bounded by the rank of the triplet $(x^*, y^*, z^*)$  in a fixed enumeration of the set and the information needed to perform the enumeration. Thus,
 \[
 \begin{array}{ll}
 C(x^* y^* \mid z^*) & \leq t_1 + t_2 + 3 - c' \log m + 2 \log(m -c'\log m) + 2 \log t_1 + 2 \log t_2 +O(1) \\
 & \leq t_1 + t_2 - (c'-2) \log m + 2 \log t_1 + 2 \log t_2 +O(1).
 \end{array}
 \]
 
On the  other hand, by the conditional version of the Symmetry of Information Equation~(\ref{e:symmetry}), there exists a constant $d$ such that for all strings $u,v,w$,
$C(uv \mid w) \geq C(v \mid w) + C(u \mid uw) - d (\log |uv|)$. It follows that
\[
\begin{array}{ll}
C(x^* y^* \mid z^*) & \geq C(y^* \mid z^*) + C(x^* \mid y^* z^*) - d \log n -O(1) \\
& \geq t_2 + t_1 - (c+2) (3 \log n) - d \log n - O(1)\\
& = t_1 + t_2 - (3c + d + 6) \log n - O(1).
\end{array}
\]
For the second inequality we have used Lemma~\ref{l:threeindep}. Note that $t_1 < n + O(1)$ and $t_2 < n + O(1)$ and $m = \sigma_1 n$. Combining the above inequalities, we obtain
$(c'-2) \log \sigma_1 n \leq (3c+d+ 10) \log n +O(1)$.
Since $c' = 3c + d + 13$, we have obtained a contradiction.~\QED
 
\section{Acknowledgments}
We are grateful to Andr\'{e} Nies and Frank Stephan for their insightful comments. In particular, Definition~\ref{d:strongindep} has emerged after several discussions with Andr\'{e}, and Proposition~\ref{p:ce} and Proposition~\ref{p:createdep} are due to Frank \cite{frank}. We also thank Jan Reimann for his assistance with establishing Fact~\ref{f:NiesReimann}.

\begin{thebibliography}{DHNT06}

\bibitem[BDS07]{bie-dot-ste:c:haussdimension}
L.~Bienvenu, D.~Doty, and F.~Stephan.
\newblock Constructive dimension and weak truth-table degrees.
\newblock In {\em Computation and Logic in the Real World - Third Conference of
  Computability in {E}urope}. Springer-Verlag {\it Lecture Notes in Computer
  Science \#4497}, 2007.
\newblock To Appear. Available as Technical Report arXiv:cs/0701089 ar
  arxiv.org.


\bibitem[C02]{Cris} C.~S.~Calude. 
 {\em Information and Randomness: An Algorithmic
Perspective}, 
 2nd Edition,
Revised and Extended, Springer-Verlag, Berlin,  2002.

\bibitem[CST06]{cal-sta-ter:j:partialrand}
C.~Calude, L.~Staiger, and S.~Terwijn,
\newblock On partial randomness.
\newblock {\em Annals of Pure and Applied Logic}, 138:20--30, 2006.

\bibitem[Cha75]{cha:j:ait}
G.~Chaitin.
\newblock A theory of program size formally identical to information theory,
\newblock {\em Journal of the ACM}, 22:329--340, 1975.



\bibitem[Cha82]{Cha82}
G.~Chaitin. G\"odel's Theorem and Information, {\em 
International Journal of Theoretical Physics} 21: 941--954, 1982.


\bibitem[DH]{dow-hir:b:algrandom}
R.~Downey and D.~Hirschfeldt.
\newblock {\em Algorithmic randomness and complexity}.
\newblock To be published by {S}pringer {V}erlag.

\bibitem[DHNT06]{dow-hir-nie-ter:j:calibrating}
R.~Downey, D.~Hirschfeldt, A.~Nies, and S.~Terwijn.
\newblock Calibrating randomness,
\newblock {\em The Bulletin of Symbolic Logic}, 12(3):411--492, 2006.


\bibitem[GV04]{gru-vit:t:shankolm}
P.~Gr\"{u}nwald and P.~Vitanyi.
\newblock Shannon information and {K}olmogorov complexity, 2004.
\newblock CORR Technical report arxiv:cs.IT/0410002, revised May 2006.

\bibitem[Kau03]{kau:t:kolmog}
S.M. Kautz.
\newblock Independence properties of algorithmically random sequences, 2003.
\newblock CORR Technical Report arXiv:cs/0301013.


\bibitem[Lev84]{lev:j:mutualinformation}
L.~Levin.
\newblock Randomness conservation inequalities: information and independence in
  mathematical theories.
\newblock {\em Information and Control}, 61(1), 1984.


\bibitem[Lut03]{lut:j:dimension}
J.~Lutz.
\newblock The dimensions of individual strings and sequences, 
\newblock {\em Information and Control}, 187:49--79, 2003.

\bibitem[May02]{may:j:dimension-kol}
E.~Mayordomo.
\newblock A {K}olmogorov complexity characterization of constructive
  {H}ausdorff dimension,
\newblock {\em Information Processing Letters}, 84:1--3, 2002.

\bibitem[ML66]{mar-lof:j:mltests}
P.~Martin-L\"{o}f.
\newblock The definition of random sequences,
\newblock {\em Information and Control}, 9:602--619, 1966.

\bibitem[Nie05]{nie:j:ktrivial}
A.~Nies.
\newblock Lowness properties and randomness,
\newblock {\em Advances in Mathematics}, 197:274--305, 2005.

\bibitem[NR06]{nie-rei:c:wtt-Kolm-increase}
A.~Nies and J.~Reimann.
\newblock A lower cone in the wtt degrees of non-integral effective dimension,
\newblock In {\em Proceedings of {IMS} workshop on Computational Prospects of
  Infinity}, Singapore, 2006.
\newblock To appear.

\bibitem[Rei04]{rei:t:thesis}
J.~Reimann.
\newblock Computability and fractal dimension,
\newblock Technical report, Universit\"{a}t Heidelberg, 2004.
\newblock Ph.D. thesis.

\bibitem[Rya84]{rya:j:dimension}
B.~Ryabko.
\newblock Coding of combinatorial sources and {H}ausdorff dimension,
\newblock {\em Doklady Akademii Nauk SSR}, 277:1066--1070, 1984.

\bibitem[Sol75]{sol:t:kolmog}
R.~Solovay.
\newblock Draft of a paper (or series of papers) on {C}haitin's work, 1975.
\newblock unpublished manuscript, {IBM} {T}homas {J}. {W}atson {R}eserach
  {C}enter, 215 pp.

\bibitem[Sta93]  {sta:93} L. Staiger. Kolmogorov complexity and Hausdorff dimension, {\em  Inform. and Comput.}  103 (1993) 159-194. 

\bibitem[Sta05]{sta:j:dimension}
L.~Staiger.
\newblock Constructive dimension equals {K}olmogorov complexity,
\newblock {\em Information Processing Letters}, 93:149--153, 2005.
\newblock Preliminary version: Research Report CDMTCS-210, Univ. of Auckland,
  January 2003.


\bibitem[Ste07] {frank} F.~Stephan. Email communication, May 2007.

\bibitem[Tad02]{tad:j:partialrand}
K.~Tadaki.
\newblock A generalization of {C}haitin's halting probability $\Omega$ and
  halting self-similar sets,
\newblock {\em Hokkaido Math. J.}, 31:219--253, 2002.

\bibitem[vL90]{vlam:j:randomness}
M.~van Lambalgen.
\newblock The axiomatization of randomness,
\newblock {\em The Journal of Symbolic Logic}, 55:1143--1167, 1990.

\bibitem[Zam90]{zam:t:kolmog}
D.~Zambella.
\newblock On sequences with simple initial segments, 1990.
\newblock ILLC Technical Report ML 1990-05, University of Amsterdam.

\bibitem[Zim07]{zim:t:extractKolm}
M.~Zimand.
\newblock Two sources are better than one for increasing the {K}olmogorov
  complexity of infinite sequences, 2007.
\newblock CORR Techical Report. arXiv:0705.4658.

\bibitem[ZL70]{zvo-lev:j:kol}
A.~Zvonkin and L.~Levin.
\newblock The complexity of finite objects and the development of the concepts
  of information and randomness by means of the theory of algorithms,
\newblock {\em Russian Mathematical Surveys}, 25(6):83--124, 1970.

\end{thebibliography}

\end{document}